\documentclass[conference, letter]{IEEEtran}
\usepackage[dvips]{graphicx}
\usepackage{cite}
\usepackage{color}
\usepackage[cmex10]{amsmath}
\usepackage{amssymb}
\usepackage{amsmath}
\usepackage{booktabs}
\usepackage{multirow}
\usepackage{dcolumn}
\usepackage{amsthm}
\usepackage{enumerate}
\usepackage{algorithmicx}
\usepackage{array}
\usepackage{mdwmath}
\usepackage{mdwtab}
\usepackage{eqparbox}
\usepackage{verbatim}
\usepackage{bm}
\usepackage[center]{caption}

\usepackage[utf8]{inputenc}
\usepackage{url}
\usepackage{enumitem}

\usepackage[]{footmisc}

\usepackage{algorithm}
\usepackage{algpseudocode}
\usepackage{color,soul}

\newtheorem{proposition}{Proposition}

\newtheorem{remark}{Remark}
\newtheorem{scenario}{Scenario}

\renewcommand{\qed}{$\blacksquare$}

\IEEEoverridecommandlockouts

\def \it {\textit}
\def \tbf {\textbf}
\def \mtc {\mathcal}

\def \ba {\begin{array}}
\def \ea {\end{array}}
\def \benu {\begin{enumerate}}
\def \eenu {\end{enumerate}}
\def \bdes {\begin{description}}
\def \edes {\end{description}}

\def \bitem {\begin{itemize}}
\def \eitem {\end{itemize}}

\def \bfl {\begin{flushleft}}
\def \efl {\end{flushleft}}
\def \bfr {\begin{flushright}}
\def \efr {\end{flushright}}

\def \beq {\begin{equation}}
\def \eeq {\end{equation}}
\def \bqa {\begin{eqnarray}}
\def \eqa {\end{eqnarray}}
\def \bqa* {\begin{eqnarray*}}
\def \eqa* {\end{eqnarray*}}

\def \bal {\begin{align}}
\def \eal {\end{align}}

\def \acro {\item}
\graphicspath{{figs/}}

\makeatletter
\newcommand*\titleheader[1]{\gdef\@titleheader{#1}}
\AtBeginDocument{%
  \let\st@red@title\@title
  \def\@title{%
    \bgroup\normalfont\large\centering\@titleheader\par\egroup
    \vskip1.08em\st@red@title}
}
\makeatother
\titleheader{\fontsize{9.5}{\baselineskip}\selectfont{This work will be published in IEEE Future Networks World Forum 2022 conference proceedings.}}
\title{On Securing MAC Layer Broadcast Signals Against Covert Channel Exploitation in 5G, 6G \& Beyond}

\begin{document}
\author{
Reza Soosahabi$^{1,2}$, Magdy Bayoumi$^{2}$\\
$^{1}$ ATI Research Center, Keysight Technologies Inc., Austin, TX 78731\\
$^{2}$ Department of Electrical \& Computer Engineering, University of Louisiana, Lafayette, LA 70503\\
Email: \{reza.soosahabi@keysight.com\}, \{magdy.bayoumi@louisiana.edu\}
}

\maketitle

\begin{abstract}
In this work, we propose a novel framework to identify and mitigate a recently disclosed covert channel scheme exploiting unprotected broadcast messages in cellular MAC layer protocols.
Examples of covert channel are used in data exfiltration, remote command-and-control (CnC) and espionage.
Responsibly disclosed to GSMA (CVD-2021-0045), the SPARROW covert channel scheme exploits the downlink power of LTE/5G base-stations that broadcast contention resolution identity (CRI) from any anonymous device according to the 3GPP standards. 
Thus, the SPARROW devices can covertly relay short messages across long-distance which can be potentially harmful to critical infrastructure. 
The SPARROW schemes can also complement the solutions for long-range M2M applications.
This work investigates the security vs. performance trade-off in CRI-based contention resolution mechanisms. 
Then it offers a rigorously designed method to randomly obfuscate CRI broadcast in future 5G/6G standards.
Compared to CRI length reduction, the proposed method achieves considerable protection against SPARROW exploitation with less impact on the random-access performance as shown in the numerical results.     

\end{abstract}


\IEEEpeerreviewmaketitle
\section{Introduction}
Covert channel schemes, in the broadest sense, are used in a wide array of security threats, such as data exfiltration, remote command-and-control (CnC) and espionage. 
The parties establishing covert channels strive to stay anonymous and circumvent security and lawful-interception systems that actively inspect the incumbent means of communication. The rapid adoption of converged connectivity solutions, such as 5G, has made covert channel schemes an integral part of most advanced security threats targeting critical industries \cite{apt2}. 
Considering the targeted layer in the OSI reference model, most of the existing covert channel schemes can be split into two categories:  exploiting layer 3 to 7 protocols and designing new PHY layer solutions (mostly wireless).

\begin{figure}[ht] \centering
    \includegraphics[width=8cm]{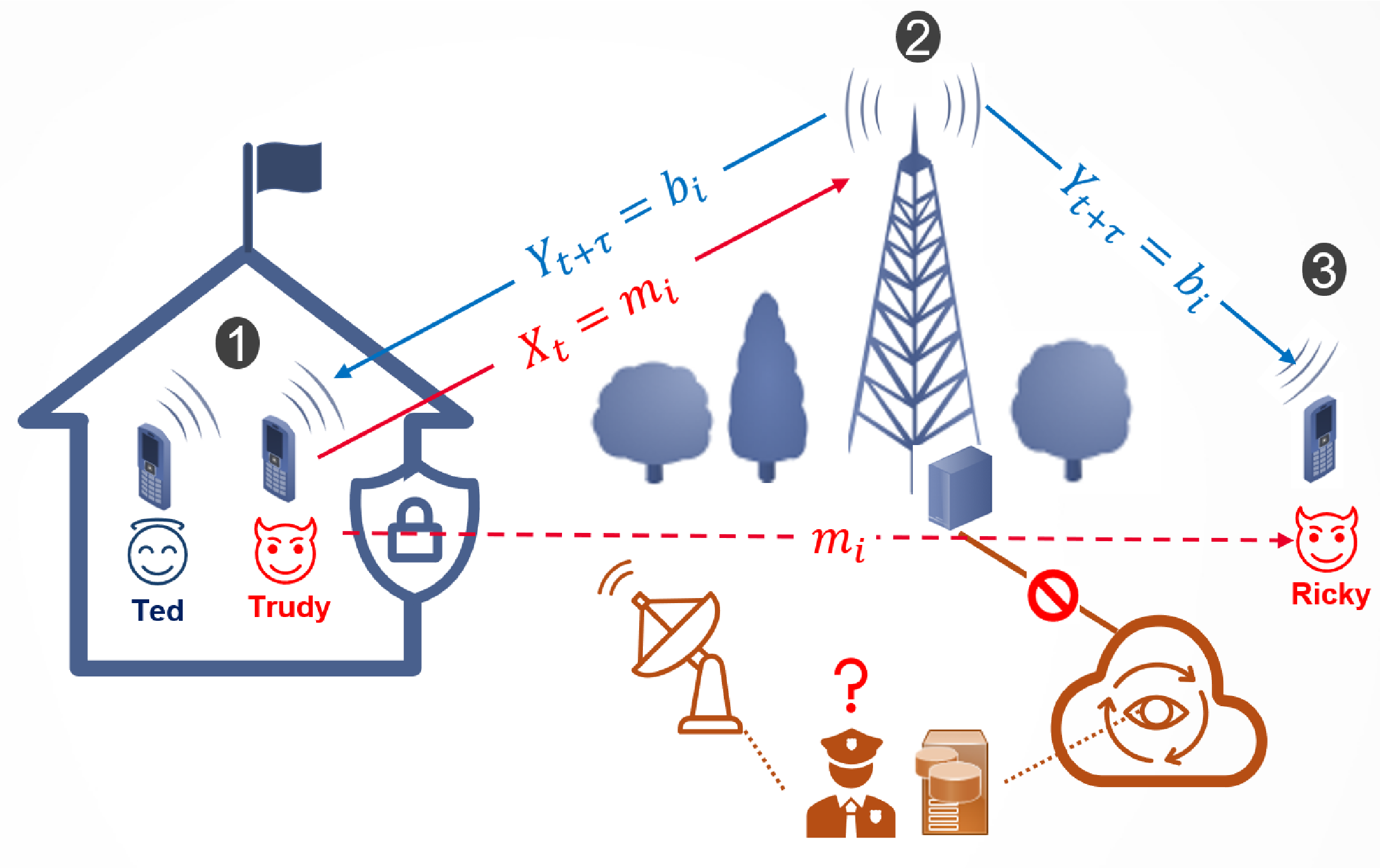}
		\caption{SPARROW exploitation model in Scenario \ref{s1}.}
    \label{model-fig}
\end{figure} 

The first category includes data exfiltration and CnC techniques that are well-known topics in the cybersecurity community. Data exfiltration involves covertly extracting and communicating sensitive information from a compromised system. Exemplary CnC implementations include malicious software that are configured to covertly communicate across the Internet. These techniques usually tunnel messages between two hosts connected to the Internet, such as ICMP and DNS tunneling \cite{cc-survey}. To counter such threats, the cybersecurity industry constantly monitors emerging techniques and adopts countermeasures to detect and block them. 

The second category includes covert communication schemes and has long been of research interest, particularly in the field of military communications \cite{military}. 
Covert communication devices usually exploit the radio spectrum without a license acquisition and generally employ low-power, ad-hoc radios that use PHY-layer technologies such as spread-spectrum. Low-power commercialized ad-hoc technologies such as LoRA and ham radios can be engineered for covert communication, but unlike commercial radios, these systems usually sacrifice transmit power and data-rate in favor of defeating spectrum monitoring and jamming systems\cite{covertlora2}. These power and data-rate limitations, along with a lack of access to elevated antennas or high transmission power, significantly reduces the operation range of these devices, particularly in indoor-to-outdoor communication scenarios\cite{lora-range2}. 

Drawing upon elements of these previous approaches, we have discovered a novel framework to harness the broadcast power of wireless macro infrastructure for covert communication \cite{defcontalk2}. 
Discussed in Section \ref{model-sec}, there are vulnerable MAC layer procedures allowing anonymous devices to trigger broadcast control signals from macro base-stations that contain covert messages. 
In Section \ref{sparrow-sec}, we detail the SPARROW covert channel scheme that exploits the contention resolution (CR) broadcast message in the random-access (RA) procedure common in LTE/5G MAC layer protocol. 
This vulnerability has been responsibly disclosed in GSMA vulnerability disclosure program under the code-name CVD-2021-0045 \cite{cvd}, and its impact on worldwide LTE/5G networks have been presented in \cite{darkreading}. 
The SPARROW schemes have the edge over existing covert channel techniques in terms of: maximum anonymity, higher operational range with low hardware footprint. 
The latter is the result of harnessing the broadcast transmit power from the macro base-stations. 
The range and impact of SPARROW schemes can be drastically amplified if discovered in cellular satellite protocols, such as GMR-2 and 5G-NTN. 
The SPARROW scheme in LTE/5G can also be used for connection-less M2M communication, where it can complement other solutions such as \cite{jover2015}.
Despite offering modest throughput, the SPARROW scheme can bypass all current security and lawful-interception systems, as well as the existing signal intelligence systems that are designed only for the PHY layer schemes.
It can 
This enables SPARROW schemes to be used in a wide variety of covert channel scenarios targeting critical infrastructure and espionage. 
Many wireless MAC protocols contain a CR procedure and the vulnerable one in 5G/LTE has been implemented in the standards for over a decade. 
This fact was the primary motivations for developing a rigorous remediation framework capable of hardening similar procedures against the SPARROW threats. 
Section \ref{entropy-sec} lays out the mathematical foundation for this novel framework called \textit{entropy-leveraging} that employs randomized obfuscation of broadcast signal during the CR procedure. 
There we also highlight a theoretical trade-off between the protection and the CR performance. 
Section \ref{elisha-sec} proposes an example of the entropy-leveraging scheme called ELISHA (entropy-leveraged irreversible salted hashing algorithm), which is used to efficiently disrupt most advanced SPARROW attacks with minimal impact on the CR performance for other users. 
The numerical results presented in Section \ref{num-sec} illustrate how to optimize the design parameters in an ELISHA remediation scheme.  
Finally, the concluding remarks are presented in Section \ref{conc-sec} that is followed by a glossary of frequently used acronyms in Appendix \ref{acros}.

\section{General Exploitation Model} \label{model-sec}
From the broadest point of view, a macro Radio Access Network (RAN) consists of a network of high-power radio access nodes (e.g. gNBs in 5G) that operate in a licensed frequency band and provide secure wireless connectivity to user devices (UEs) across a large geographic area. 
The UEs have to authenticate with a core network (CN) entity before accessing any of the network services. 
There are also CN servers that collect user activity metadata such as service usage and user location. 
The metadata are then used internally for network optimization and shared with government authorities in compliance with Lawful Intercept (LI) regulations \cite{LI2}. 

Unlike their ad-hoc counterparts, macro RANs implement centralized MAC layer protocols that prohibit the UEs from untraceable peer-to-peer (P2P) wireless communications. 
Thus, we provide a novel framework to identify potential weaknesses in the MAC layer protocol procedures that enable anonymous UEs to exploit the macro radio access nodes for long-range P2P communication. For the ease of illustration, we have formulated the following hypothetical exploitation scenario in a terrestrial cellular context. It can be easily extended to other wireless technologies such as non-terrestrial networks:  

\begin{scenario}\label{s1}
Trudy intrudes a cybersecurity air-gaped facility under heavy surveillance and wishes to covertly send a set of messages to her counterpart Ricky with a passive receiver outside. 
They cannot access the incumbent network. They also cannot leverage ad-hoc radios due to spectrum surveillance and insufficient signal range. 
However, both are equipped with low-power programmable UEs that can interact with the same nearby macro 5G base-station. 
Knowing a vulnerability in 5G MAC layer protocol, Trudy programs her UE to exploit the nearby gNB to broadcast (relay) her messages to Ricky without authenticating with a carrier network.
\end{scenario}

MAC layer protocol procedures can be expressed as a flow of messages exchanged between each UE and a base-station. 
In this case, Trudy and Ricky construct a \it{code-book} for their communication scheme. 
It defines two sets of $N$ messages $\mathcal{M} = \{m_1,m_2,\cdots,m_{M}\}$ and $\mathcal{B} = \{b_1,b_2,\cdots,b_{M}\}$ that respectively denote Trudy's possible unlink transmissions and their resulted broadcast messages from the base-station.
Let the random variable $X_t\in\mathcal{M}$ stand for Trudy's message sent at time slot $t$. 
The base-station response to $X_t$ can be modeled with another random variable $Y_{t+\tau}\in\mathcal{B}$, where $\tau$ is the time lapse from Trudy's transmission moment until the base-station broadcasts the response message. 

\begin{proposition}\label{p1}
A centralized wireless MAC layer protocol is deemed vulnerable to exploitation Scenario \ref{s1} if any of its procedures allows forming $\mathcal{M}$ and $\mathcal{B}$ sets of, respectively, uplink and downlink broadcast messages that satisfying the following conditions:
\begin{enumerate}
	\item \tbf{Passive Reception:} Any passively scanning device within the radio coverage area can decode the broadcast messages in $\mathcal{B}$ without a connection establishment or exchange of PHY channel state information. 
	\item \tbf{Bijectivity:} For $1<i\leq M$, Receiving downlink message $b_i$ at time $t+\tau$ is almost surely the result of sending $m_i$ in uplink at time $t$, i.e. 
	\begin{align*}
		Pr(Y_{t+\tau} = b_i | X_t = m_i) \approx 1
	\end{align*}
	\item \tbf{Anonymous Uplink:} Sending messages in $\mathcal{M}$ does not need authentication.
	\item \tbf{Stateless Uplink:} The uplink messages in $\mathcal{M}$ can be independently transmitted in consecutive time slots,
	\begin{align*}
		Pr(X_t = m_i | X_{t-\Delta t} = m_j) = Pr(X_t = m_i)~,~ \Delta t > \tau.
	\end{align*}
\end{enumerate}
\end{proposition}

Illustrated in Fig. \ref{model-fig}, the bijectivity condition enables Ricky to learn about $m_i$, from Trudy, once it successfully decodes its corresponding broadcast message $b_i$ from the base-station. 
Hence, Trudy can anonymously exploit the base-station transmit power to deliver her message to Ricky while bypassing incumbent surveillance/LI systems. 
The stateless uplink condition allows Trudy to send consecutive messages in every $\tau$ seconds that translates to maximum data rate in the order of $\frac{\log M}{\tau}$ bits per seconds. 
Nevertheless, the poor channel condition and the similar broadcast messages intended for the other UEs in the cell can mitigate Ricky's ability to decode Trudy's messages. 
Therefore, $\mathcal{B}$ has to be a subset of all possible broadcast messages to reduce the impact of other UE activity in the cell. 
The broadcast resource control messages in most MAC layer protocols are potential candidates to construct $\mathcal{B}$ since they can better survive poor channel condition and reach father distances. 
Other practical approaches to reduce communication errors may include message repetition or establishing a similar reverse link from Ricky to Trudy for acknowledgments and synchronization. 
\begin{figure}[htb] \centering
    \includegraphics[width=8cm]{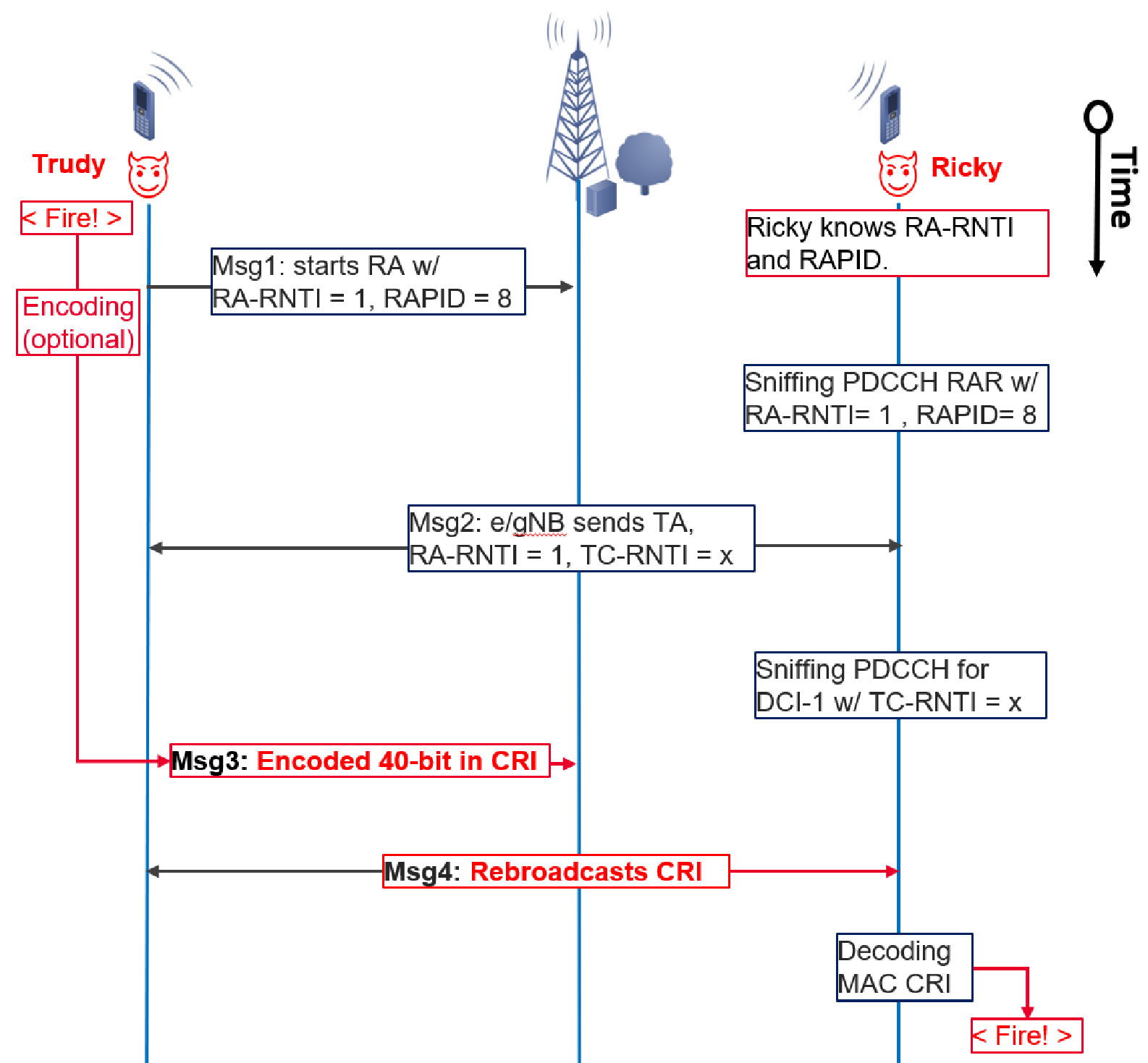}
		\caption{Exploiting RA procedure in LTE/5G.}
    \label{sparrow-fig}
\end{figure}
\section{SPARROW Exploitation Scheme in LTE \& 5G } \label{sparrow-sec} 
A quintessential instance of the weakness described in Proposition \ref{p1} has been discovered in the random-access (RA) contention resolution (CR) procedure described in 3GPP standard specification TS36.321 \cite[sec. 5.1.5]{ts36}). 
This procedure dates back to early cellular standards and likely to be present in other wireless standards utilizing centralized MAC coordination functions. 
Adopting cellular terminology, a base-station has to break up the contention (race) condition between UEs simultaneously establishing connection using the same random access unlink resource. The practical implications of this vulnerability have been presented in \cite{defcontalk2, darkreading}.

UEs establishing RRC connection with a gNB lack any prior C-RNTI assignment, which serves as a ephemeral MAC address for resource management among the UEs. 
Thus, the standard requires each UE to randomly select a $40$-bits Contention Resolution Identity (CRI) in the RRC Connection request ($Msg3$). 
Oblivious to others, each UE compares its CRI to what it receives in gNB $Msg4$ broadcast (using simplest coding modulation scheme). 
It proceeds with RA if they match, otherwise it backs off for some random duration before retrying RA \cite[sec. 6.1.3.3-4]{ts36}.
This procedure meets all of Proposition \ref{p1} criteria with $\mtc B = \mtc M$.

Illustrated in Fig. \ref{sparrow-fig}, Trudy encodes her message in $Msg3$ CRI and anonymously send it to the victim macro gNB. 
Upon receiving her $Msg3$, the victim gNB broadcasts the same message in $Msg4$ for Ricky to decode. 
Ricky can passively decode all CRI broadcast from the victim gNB, or limit its search space by having a prior agreement with Trudy about her RAPID and RA-RNTI. 
If so, Ricky only decodes the DCI values associated with the expected RA-RNTI for $Msg2$. Upon receiving a matching $Msg2$, it extracts its TC-RNTI content to detect and decode the subsequent $Msg4$ and check its content against the code-book $\mtc B$.

Trudy can break longer messages into chunks of 40-bits (or less) and transmit them in consecutive attempts\protect\footnote{A proof-of-concept has been implemented and verified using Keysight UXM5G$^{\circledR}$ and UeSIM$^{\circledR}$. Its video recording has been responsibly disclosed to GSMA and included in the presentation at \cite{defcontalk2}.}. There has been studies on the average RA-procedure duration (from $Msg1$ to $Msg4$) including \cite{rach-time2} expecting it to be around $30 ms$ in typical LTE deployments. Taking this estimate and accounting for additional $10 ms$ of back-off between multiple attempts, Ricky and Trudy can achieve near $1 kbps$ throughput in this scheme. The offered throughput suits IoT and M2M applications that currently use low-power technologies such as LoRA \cite{lora-range2}. However, SPARROW scheme can achieve longer range in cluttered environment without any direct access to RF spectrum.

It will be appreciated that the RA procedure is agnostic to the PHY layer frequency band. However, the lower frequency bands in LTE/5G RAN better suits the objectives of Scenario \ref{s1}. As far as RA concerned, the cell range depends on the PRACH preamble zero-correlation-zone configuration (Ncs) of the gNB (illustrated in section 24.8 of \cite{bullets2}). For typical outdoor LTE macro cells, Ncs is set to 9 or larger values that enables UEs to perform RA as far as 5 miles from the cell. 5G-NR (new radio) standards enable utilizing higher frequency bands above $6$ GHz (FR2) that rely on beam-forming and multiple-antenna transmission modes. Nevertheless, the underlying RA procedure PHY layer is still very similar to LTE in sub-$6$ Ghz (FR1) and therefore, more promising.

\begin{figure}[htb] \centering
    \includegraphics[width=8cm]{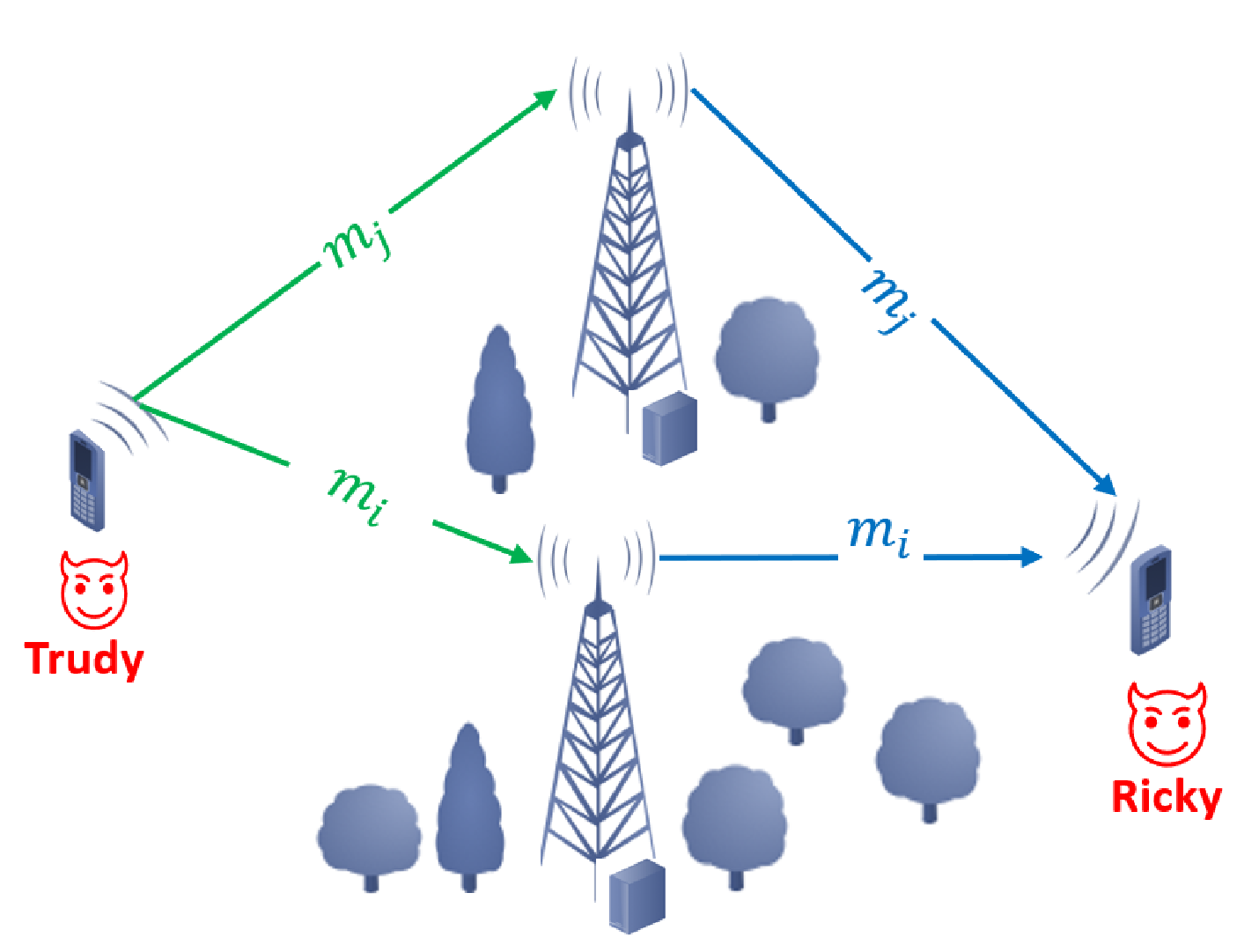}
		\caption{Exploiting overlapping cells (sectors) for throughput enhancement.}
    \label{parallel-fig}
\end{figure}

Depending on the application, SPARROW UEs (Ricky and Trudy) can exploit multiple LTE/5G carriers for throughput or operational range enhancements. Figure \ref{parallel-fig} shows how two cells can be exploited to achieve parallel covert channels. With the exception of very rural environments, UEs within the range of a few miles can be covered by multiple overlapping LTE or 5G sectors with multiple carrier frequencies, which can be simultaneously exploited for more throughput or a reverse link from Ricky to Trudy. 
\begin{figure}[tb] \centering
    \includegraphics[width=8cm]{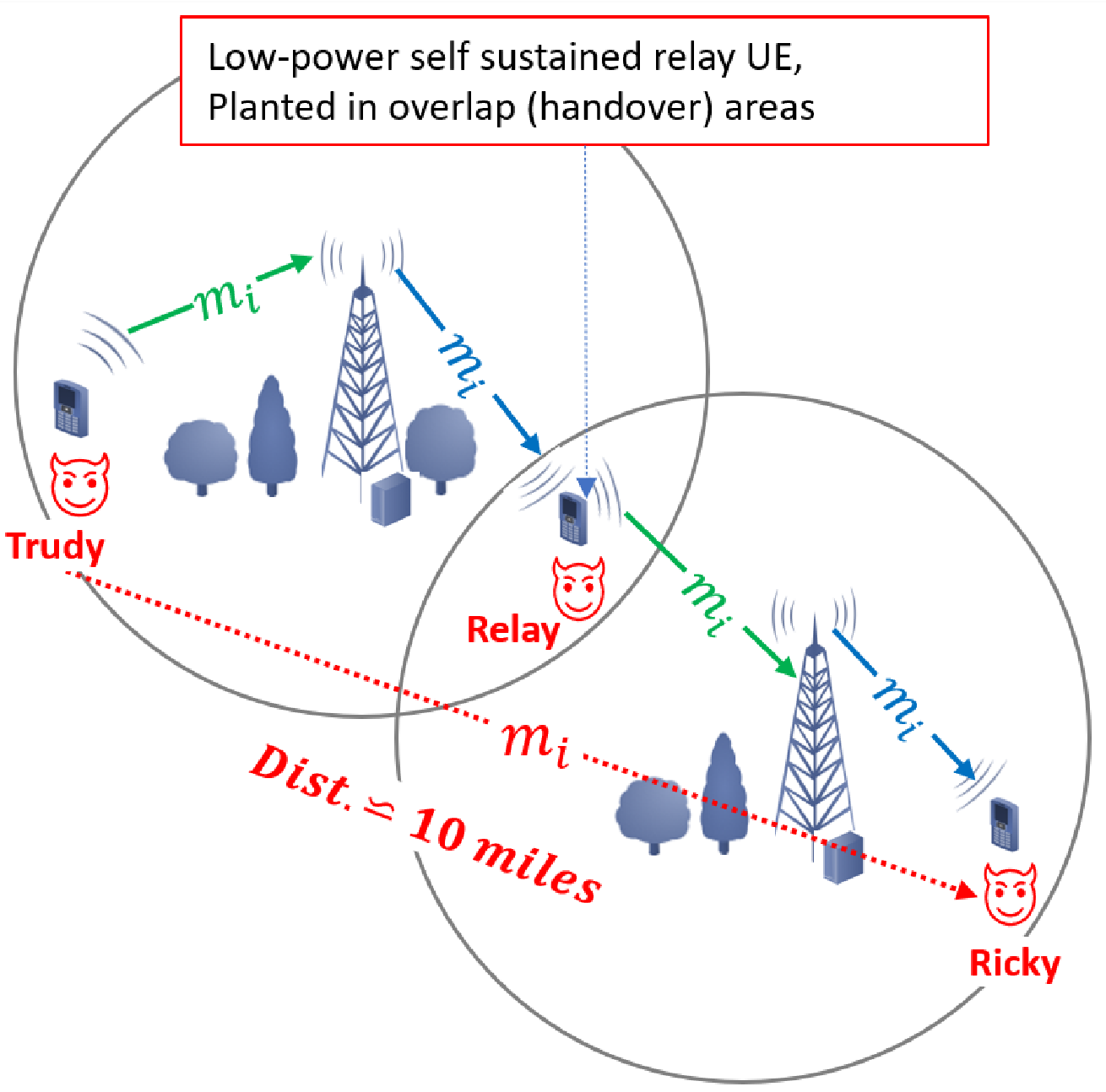}
		\caption{Exploiting adjacent cells (sectors) for range enhancement with a relay.}
    \label{relay-fig}
\end{figure}
Figure \ref{relay-fig} depicts a more interesting case involving a relay UE to extend the operational range beyond a single cell coverage. Relay UEs are placed in the handover (coverage overlap) region between adjacent cells. These relays can be configured to act as a proxy for Ricky, receiving a message in one cell and transmitting it in another adjacent cell. The SPARROW UEs are effectively low-power cellular modems that can operate off of rechargeable batteries. Thus a rechargeable relay UE can operate from any inconspicuous location in between cells. One can possibly create a wide-area IoT mesh using relay UEs communicating via SPARROW.

\section{Analysis of Entropy-Leveraging Framework} \label{entropy-sec}
Averting drastic changes to the existing CR procedure, the future 5G/6G MAC protocols can block SPARROW covert channel by reducing its maximum achievable bit rate (channel capacity). 
Reducing the size of CRI is a simple approach that likewise proportionally decreases the efficiency of the CR for the ordinary users. 
Here we present a framework for the methods that derive $Msg4$ via applying entropy (random obfuscation) operations on the $Msg3$ received at the gNB. 
\subsection{Formulation}
Expanding the notation in Section \ref{model-sec} and Scenario \ref{s1}, the CR procedure in the proposed framework occurs in the following steps: 
\begin{enumerate}
\item \tbf{Uplink Message:} 
Let $X_i \sim \mathcal{U}(2^{-N})$ denote the $N$-bits discrete random variable denoting $Msg3$ transmission by the $i$-th UE contending for the same PRACH resource in the cell. Analyzing a single exchange, the time has been omitted for brevity. 
The random variable $X'$ denotes Trudy's $Msg3$ transmission from code-book $\mathcal{M} \subset U_N$.

\item \tbf{Obfuscated Broadcast:} The sensitivity of PRACH preambles to the channel timing ensures that the gNB only receives (at random) one of the $Msg3$ transmissions, denoted by $X \in \{X_1,X_2,X'\}$. 
It then derives $Y = [B(X), h]$ for transmission in $Msg4$ broadcast, where $B$ is the \it{broadcast random obfuscation function} along with the hint value $h$. 
The visiting UEs should be made aware of the obfuscation function $B$ via announcements in the periodic SIB messages or in $Msg2$. 
The value of $h$ assists the intended UE, whose $Msg3$ received at gNB, to make correct RA decision and the rest of them to back off.

\item \tbf {UE Decision:} Any choice for $B$ should be accompanied with a well-defined UE decision function $D = D(Y,X_i) \in \{0,1\}$, where $0$ and $1$ are, respectively, interpreted as RA success and failure commands for the $i$-th UE. 
Resolving the contention requires that only one of the contending UEs to arrive at $D=1$. 
The decision function should also eliminate the possibility of a live-lock situation where all UEs deduce RA failure.
\begin{align}
Pr(D(Y, X_i) = 1 | X = X_i) = 1 
\end{align}
\end{enumerate}
Knowing the choice for $B$ and code-book $\mtc{M}$, Ricky attempts to optimally decode $X'$ from $Msg4$ with minimum errors. 
Let $X'' = E(Y)$ be the random variable representing Ricky's estimated codeword. 
Ricky designs $E(Y)$ to minimize its estimation error probability, $Pr(X'' \neq X')$. 

An effective CR procedure requires low \it{CRI collision probability}, denoted by $P_C$, to ensure only one of the contending UEs succeeds in RA. 
In practice, having more than two UEs simultaneously attempting RA using the same preamble is a rare event that may occur in occasions such as a base-station recovery. 
Henceforth, we only consider the contention scenario between two UEs, $i\in{1,2}$ whose probability is 
\begin{align} \label{pcdef}
P_{C} := Pr(D(Y, X_2) = 1 | X = X_1).
\end{align}
Here we assume both UEs can decode $Msg4$ error free. 
Considering the uniform i.i.d property of $X_i$, the expression in \eqref{pcdef} can be further expanded to 
\begin{align} \label{pcexpanded}
P_{C} = \mathop{\sum\sum}\limits_{\substack{i,j\in U_N \\ i \neq j}}\frac{Pr(D(Y,i)=1|X_2=i, X=j)}{2^{2N}} + \frac{1}{2^{N}}.
\end{align}
It implies that $2^{-N}$ is the minimum achievable value for $P_C$ when $B(X)$ is a deterministic bijective function, such as $B(X) = X$ and $D(X_i,Y) = \delta(X - X_i)$ described in the current standard versions.

\subsection{Key Trade-Off}\label{tradeoff}
The data rate of SPARROW UEs depends on how they design the code-book $\mtc M$ and the estimation function $E(Y)$ to overcome the effects of the entropy operation $B(X)$. 
Analyzing the channel capacity for the SPARROW UEs can reveal the inherent trade-off between the protection and performance in this framework. 

Given $X = X'$, the perfect code-book $\mtc M$ maximizes the channel capacity for the SPARROW UEs, defined as the maximum number of error-free bits they can transmit in each attempt. This condition requires maximizing the following mutual information:
\begin{align} \label{mutual}
I(X;Y) = H(X) - H(X|Y), 
\end{align}
where $H(.)$ denotes Shannon entropy. 
The remediation methods in the entropy-leveraging should design $B(X)$ and $D(Y,X)$ so that $H(X|Y)$ is maximized. 
On the other hand, applying Fano's inequality to the expression in \eqref{pcexpanded}, it can be shown that $H(X|Y)$ directly contributes to a lower bound on $P_C$ \cite{fano2}. 
Hence, any method has to strike a balance in the trade-off between the CR performance, low $H(X|Y)$, and degrading the channel capacity for the SPARROW UEs, high $H(X|Y)$. 

Considering the practical aspects of the SPARROW exploitation, forcing the channel capacity down to a few bits per attempt can block most SPARROW exploitation scenarios. 
The SPARROW channel rate should be large enough to accommodate the overheads for synchronizing the endpoints and circumvent message confusion with other UEs in the cell \cite{defcontalk2}.

\begin{remark}\label{lastrmk}
For the same CRI collision probability, the candidate methods in the entropy-leveraging framework should achieve more reduction in the SPARROW channel capacity than the simple CRI length reduction approach. In other words, they provide a better performance / protection trade-off by preventing the SPARROW UEs from optimizing their communication code-books.
\end{remark}

\section{Entropy-Leveraged Irreversible Salted Hashing Algorithm (ELISHA)} \label{elisha-sec}
Our proposed example of entropy-leveraging method prevents the SPARROW UEs from forming code-books with forward-error-correcting (FEC) property against the randomized obfuscation of the broadcast messages in each RA attempt. 
They can no longer predict how the messages in the uplink code-book $\mtc M$ map to the broadcast messages in $\mtc B$. 
\begin{figure}[htb] \centering
    \includegraphics[width=8cm]{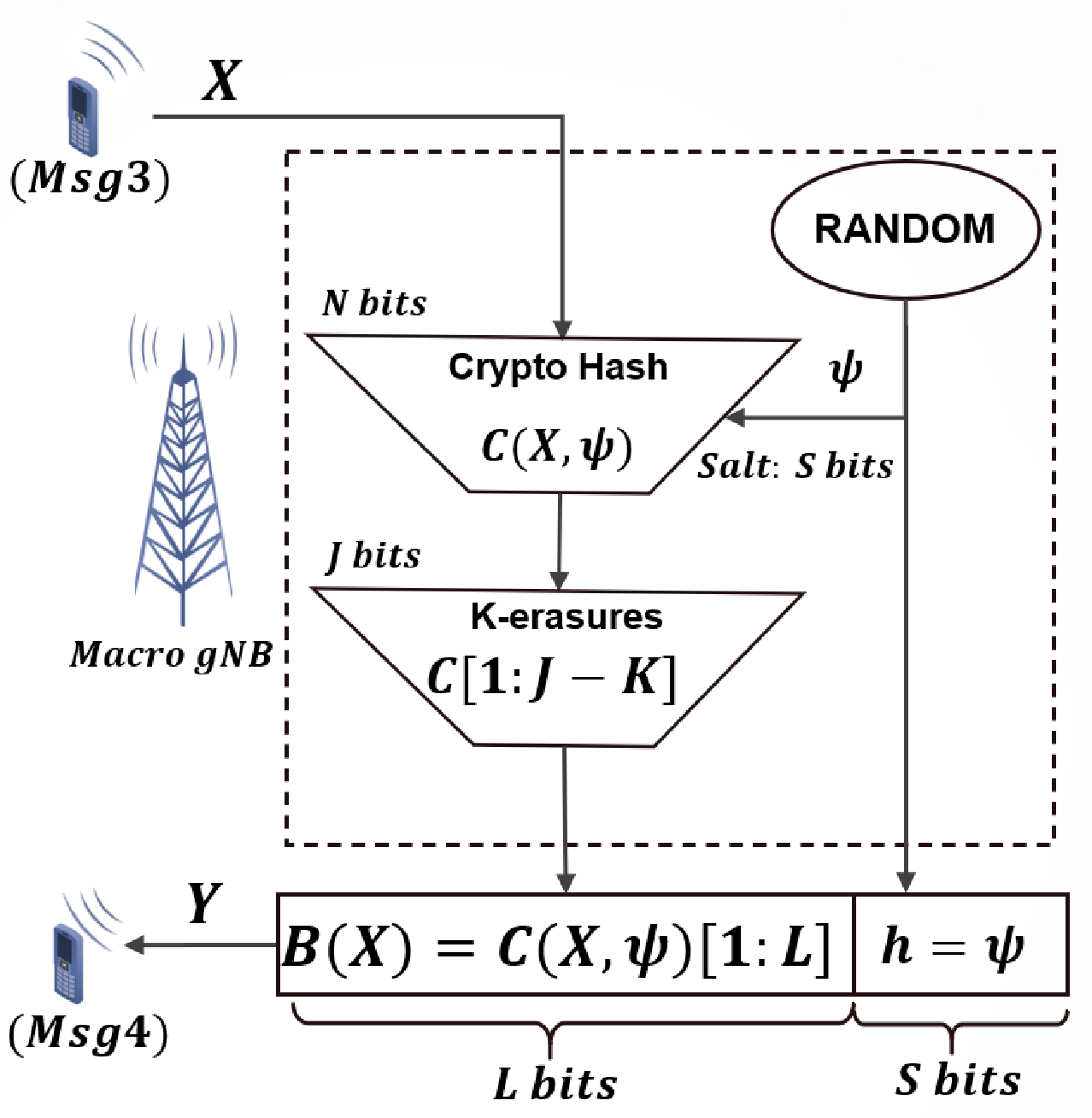}
		\caption{ELISHA transformation of $Msg3$ to $Msg4$.}
    \label{elisha-fig}
\end{figure}

Fig. \ref{elisha-fig} illustrates the components of ELISHA broadcast obfuscation function performed at the base-station. It randomly maps the $N$-bits CRI in $Msg3$ to an $L$-bits output digest in $Msg4$ in each RA attempt. 
The received $N$-bits CRI is processed by a standard cryptographic hash function (CHF), denoted by $C(X,\psi)$, with a randomly generated salting nonce $\psi$ ($S$-bits size) to produce an $J$-bits hash digest. 
Comprised of complex arithmetic modules, most CHFs deterministically map large variable-length inputs to a much smaller fixed-length outputs such that small input variations are amplified to large output variations. 
Working with a massive input space, the CHFs are computationally irreversible in the sense that the input cannot be estimated from the output. 
However, applying them to relatively short fixed-size inputs, such as CRI, dramatically weakens their irreversibility while modestly improving bijectivity (less collision). 
This concern is addressed by arithmetically mixing the input with a unique random salting byte-string $\psi$ to achieve strong irreversibility by enlarging the CHF input space\cite{chfibm}. 
There are a variety of choices for $C(X,\psi)$, ranging from sophisticated SHA family to simpler MD family that mostly result in $J > N$. 

The CHF output then undergoes a simple truncation process, where a block of $K$ bits is erased at a known position resulting in $L = J-K$ bits of output digest. 
Finally, the gNB broadcasts the following $L+S$ bits in the $Msg4$. 
\begin{align}
 Msg4:~~Y = [~B(X) = C(X,\psi)[1:L],~ h = \psi~],
\end{align} 
where the array operand $[1:L]$ represents selecting $L$ bits from hash digest $C(X,\psi)$.
Upon receiving the $Msg4$, the $i$-th contending UE computes $B(X_i)$ for its previously transmitted CRI using the same salt byte-string hinted in $Msg4$. 
Forming the following decision function, it proceeds in RA only if the computed digest value is identical to what received in $Msg4$:           
\begin{align}
 UE_i:~~ D(Y,X_i) = \delta(~C(X_i,\psi)[1:L] - B(X)~)
\end{align} 
Assuming $C(X,\psi)$ to have negligible collision rate, it can be shown that the CRI collision probability depends on the output length $L$ as follows
\begin{align}\label{elishapc}
\text{ELISHA:}~~~~P_C \approx \frac{1}{2^L}.
\end{align}

Truncating $C(X,\psi)$ output unpredictably perturbs bijectivity condition for the SPARROW UEs due to the variable salting in each RA attempt. 
As laid out in the following Proposition, this nullifies SPARROW UEs' attempts to regain bijectivity via code-book design. 
Furthermore, their maximum achievable rate can be computed based on only the code-book size, $M$, regardless of its content.

\begin{proposition} \label{p2}  
Given any choice of code-book $\mtc M=\{m_i\}_{i=1}^{M}$, ELISHA obfuscation function $C(X,\psi)[1:L]$ randomly transforms its elements to a set of $L$-bits messages $\mtc{B} = \{b_j\}_{j=1}^{\bar M}$ of a random size $\bar M \leq M$, where $b_j \sim \mathcal{U}(2^{-L})$.
Predicting the elements in $\mtc B$ is computationally infeasible due to the irreversible property of $C(X,\psi)$. 
In each attempt, Ricky can properly decode a message $m_i$ from Trudy only if it is uniquely transformed, i.e. $C(m_i,\psi)[1:L] \neq C(m_j,\psi)[1:L]$ for any $m_j \neq m_i$. 
This leads to the following formula for the maximum theoretical data rate of the SPARROW exploitation channel, denoted by $R_M$ in bits-per-attempt.
\begin{align}\label{rm}
R_M := \left(1 - \frac{1}{2^L}\right)^{M-1}\log_2{M}
\end{align}
\end{proposition}

Appendix \ref{proof} includes the proof of the derivation in \eqref{rm} using the known analysis of the random occupancy problem. Extended to a continuous form, it can be shown that $R_M$ is a concave function with a single maximum value. We can also leverage the following approximation when computing $R_M$ for large values of $2^L$.
\begin{align}\label{approxrm}
R_M \approx e^{{(1-M)}/{2^L}}\log_2{M}
\end{align}

Finally, the SPARROW UEs are left only with optimizing the code-book size $M$ such that it maximizes the theoretical capacity of the exploited channel in the long-run, denoted by 
\begin{align}\label{rmax}
R_{max} := \max_{M < L} \{R_M\}
\end{align}

It is worth reiterating that the maximum capacity in \eqref{rmax} may be practically unattainable for the SPARROW UEs in most scenarios. 
It will require successive transmissions of the same $Msg3$ until it is uniquely transformed through the obfuscated broadcast for Ricky to decode. 
Ricky needs some synchronization mechanism to know if Trudy is sending a new message or retransmitting the same one. 
This will require allocating part of the channel capacity for sequencing flags in addition to avoiding message confusion with other UEs. 
Therefore, a network operator can dismay SPARROW UEs by configuring the ELISHA parameters such that $R_{max}$ drops below a byte ($8$ bits) per attempt.

If adopted in the standard for the CR procedure, it will be up to the network operators to optimize $C(X,\psi)[1:L]$ parameters per base-station to reach the desired levels of protection and performance. 
Another advantage of ELISHA is having the derivations in \eqref{rm} and \eqref{elishapc} that can significantly streamline the optimization process. 
The operators can choose a CHF from a set of choices set forth by the standard, taking into account the performance factors such as resource allocation for the enlarged $Msg4$, CHF computational complexity and its collision rate. 
In search of the optimal $L$, they should also consider the cell loading and presence of critical facilities in the area to, respectively, choose the maximum tolerable values for $P_C$ and $R_{max}$. 
The optimal value of $L$ then dictates the number of erased bits from the CHF output, $K = J - L$. 
The value of $K$ can also be dynamically optimized based on the live traffic information, as long as the standard enables signaling $K$ in a periodic SIB broadcast or $Msg2$ to the UEs.

\section{Numerical Results} \label{num-sec}
The numerical results in this section demonstrate the variation of exploited channel capacity in \eqref{rm} with respect to the code-book size $M$ and its peak value, defined in \eqref{rmax}, for different broadcast output size $L$. 
Then this maximum value is used to evaluate the efficacy of the proposed method against the CRI length reduction method mentioned in Remark \ref{lastrmk}.

\begin{figure}[t] \centering
    \includegraphics[width=8cm]{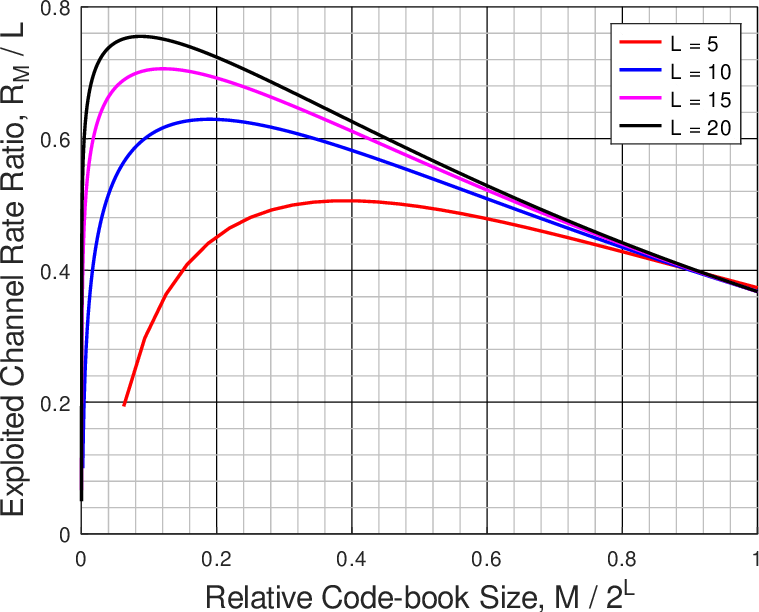}
		\caption{Impact of code-book size on the exploited channel data rate in ELISHA.}
    \label{ratem-fig}
\end{figure}

Fig. \ref{ratem-fig} plots the maximum achievable data rate $R_M$ in \eqref{rm}, normalized by its maximum value $L$, versus the code-book size, $M$. 
To compare the curves produced for different $L$, the code-book size is also normalized by its maximum value $2^L$. 
These curves indicate that $R_M$ is a concave function with respect to $M$, as expected with most channel capacity derivations. 
The rate is proportional to small values of $M$ until it reaches $R_{max}$, and then incrementing $M$ degrades the chance of having uniquely-mapped messages. 
On the other hand, increasing $L$ (reducing the truncation size $K$) apparently reverses the loss of bijectivity due to truncation that leads to larger $R_{max}$.      
\begin{figure}[htb] \centering
    \includegraphics[width=8cm]{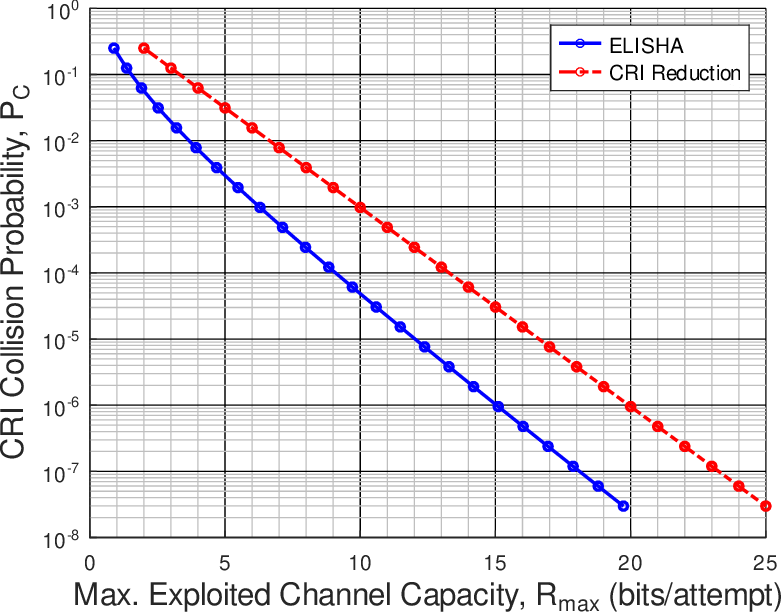}
		\caption{ELISHA beating the CRI length reduction method in the performance vs. protection trade-off.}
    \label{ratepc-fig}
\end{figure}	
Fig. \ref{ratepc-fig} demonstrates the role of output message length $L$ in balancing the key trade-off between the exploited capacity $R_{max}$ and CRI collision $P_C$ as discussed in Section \ref{tradeoff}. 
For a given $P_C$, it shows that ELISHA has achieved the framework objective in Remark \ref{lastrmk}, by achieving $20\% - 50\%$ reduction in the theoretical exploited capacity compared to the simple CRI length reduction method. 
In scenarios where low $P_C$ preferred to low $R_{max}$, ELISHA can achieve $10$ times, or more, reduction in $P_C$ compared to the CRI length reduction method for a fixed $R_{max}$ value. 
Converted to a table, the plot in Fig. \ref{ratepc-fig} can help the network operators to look up the optimal value of $L$ per scenario and configure the truncation length $K$ on base-stations accordingly.

In summary, the presented numerical results indicate that ELISHA, despite its simplicity, can be an effective solution to protect CR procedures in 5G, 6G and other wireless macro technologies against SPARROW exploitation schemes. 
It can be further improved to be considered for adoption in the future standards as a secure CR procedure option that is enabled on wireless macro infrastructure serving critical facilities sensitive to covert channel exploitation.

\section{Conclusion \& Future Works}\label{conc-sec}
This work proposed a novel framework to identify and mitigate a recently disclosed covert channel exploitation scheme in MAC layer protocol of commercial wireless technologies. 
In this framework, the SPARROW scheme uses the broadcast power of incumbent wireless networks to covertly relay messages across a long distance without requiring any authentication. 
This enables the SPARROW scheme to bypass all security and lawful-intercept systems and gain advantage over existing covert techniques in terms of: maximum anonymity, longer range and less hardware. 
This vulnerability has been in the contention resolution procedure of the LTE/5G standards for a long time. 
Hence, this work offers a remediation framework tailored for this common procedure using random obfuscation of the vulnerable broadcast messages.
A rigorously designed example in this framework, ELISHA can effectively disrupt the most sophisticated SPARROW schemes with manageable system performance overheads. 

Researchers are encouraged to investigate the SPARROW vulnerability conditions, outlined in Proposition \ref{p1}, in other wireless MAC protocols, particularly in the emerging satellite standards such as 5G-NTN. 
Exploitation of non-terrestrial wireless infrastructure can significantly amplify the operational range of SPARROW covert channels. 
The general vulnerability framework in Section \ref{model-sec} can be readily extended to covert channels schemes using other passively measurable aspects of the cell radio that can be exploited for a covert channel complementing the examples discussed in \cite{wifi} for the limited range WLAN standards.

\appendices

\section{Frequently Used Acronyms}\label{acros}
\addcontentsline{toc}{section}{Nomenclature}
\begin{IEEEdescription}[\IEEEsetlabelwidth{TC-RNTI}]
\acro[gNB] Cellular Base-Station, in 5G terminology
\acro[CHF] Cryptographic Hash Function 
\acro[CR] Contention Resolution
\acro[CRI] Contention Resolution Identity, arbitrarily selected by UEs during CR
\acro[C-RNTI] Cell RNTI, assigned by gNB to each UE
\acro[DCI] Downlink Control Information, transmitted on PDCCH enabling UEs to decode their data
\acro[GSMA] GSM Association, worldwide trade organization
\acro[LI] Lawful Intercept, in commercial communications
\acro[Msg1] Uplink random-access preamble transmission
\acro[Msg2] Downlink random-access response 
\acro[Msg3] Uplink RRC connection request containing CRI
\acro[Msg4] Downlink Contention Resolution Response
\acro[M2M] Machine-to-Machine Communication  
\acro[PDCCH] Physical Downlink Control Channel
\acro[PRACH] Physical Random-Access Channel
\acro[RA] Random-Access
\acro[RAN] Radio Access Network
\acro[RAPID] Random-Access Preamble Identifier
\acro[RA-RNTI] Random-Access RNTI
\acro[RRC] Radio Resource Control, the layer 3 protocol between UE and gNB 
\acro[RNTI] Radio Network Temporary Identifier
\acro[SIB] System Information Block, providing cell information to accessing UEs
\acro[TC-RNTI] Temporary Cell RNTI, assigned by gNB to each UE
\acro[UE] User Equipment, in cellular terminology

\end{IEEEdescription}

\section{Computing Exploited Channel Capacity}\label{proof}
Revisiting Proposition \ref{p2}, the random obfuscation in ELISHA scheme turns the exploited channel into a typical erasure channel such that the receiver cannot decode the messages mapped to the same broadcast output, as if they were erased. 
Thus, the capacity of the exploited channel is given by
\begin{align}\label{mec}
R_{M} = (1-p_e)\log_2 M,
\end{align} 
where $p_e$ is the probability of the erasure event, i.e. $1-p_e$ denotes the probability of $B(m_i)$ to be unique for a transmitted $m_i \mtc M$.
In each channel attempt, let $\mtc Q \subseteq \mtc M$ denote the uniquely obfuscated messages, i.e.
\begin{align}
\mtc Q := \{m_i \in \mtc M ~|~ \nexists~ m_j \in \mtc M : B(m_j) = B(m_i)\}.
\end{align}
Now considering that Trudy's transmitted message, $X'$, is a uniform random variable selected from $\mtc M$, the probability in \eqref{mec} is calculated as:
\begin{align}\label{1pe}
1-p_e &= \sum_{k=1}^{M} Pr(X' \in \mtc Q ~|~|Q| = k)Pr(|Q| = k)\nonumber\\
      &= \sum_{k=1}^{M} \frac{k}{M} Pr(|Q| = k)\nonumber\\
			&= \frac{E(|Q|)}{M},
\end{align}
where $E(|Q|)$ is the expected (aka the mean) size of $\mtc Q$. 

The random obfuscation in ELISHA scheme can be analyzed in the context of classic occupancy (balls-and-bins) problems \cite{parzen}, where $M$ distinct messages are randomly mapped to $2^L$ possible broadcast output messages. The calculation of $Pr(|Q| = k)$ in \eqref{1pe} involves cumbersome recursive equations investigated in \cite{tdm1999}. However, we can derive a compact expression for $E(|Q|)$ using the linear properties of expectation function. 
Let $\{b_i\}_{i=1}^{2^L}$ denote the set of all possible $L$-bits broadcast output messages, and $U_i$ be an indicator random variable such that 
\begin{equation}
U_i := 
	\begin{cases}
		1,& \mbox{if}~\exists ~m\in \mtc Q : B(m) = b_i\\
		0,& \mbox{if}~\nexists ~m\in \mtc Q : B(m) = b_i
	\end{cases}
\end{equation}
Therefore, we can calculate 
\begin{align}\label{e1q1}
E(|Q|) &= \sum_{i=1}^{2^L} E(U_i) = \sum_{i=1}^{2^L} Pr(U_i = 1)\nonumber\\
			 &= \sum_{i=1}^{2^L} \frac{M(2^L - 1)^{M-1}}{2^{LM}}\nonumber\\
			 &= M\left(1-\frac{1}{2^L}\right)^{M-1}.
\end{align}
Substituting the result of \eqref{e1q1} in \eqref{1pe} and then \eqref{mec}, we obtain the compact formula in \eqref{rm} for $R_M$. ~~~~~~~~~~~~~~~~~~~~~~~~~~~\qed

\section*{Acknowledgments} \label{ack}
We would like to thank these individuals at Keysight Technologies Inc. who assisted with disclosure and presentation of this work:
Chuck McAuley for presentation at \cite{defcontalk2}, Befekadu Mengesha and Lucal Mapelli for PoC implementation in \cite{defcontalk2}\cite{cvd}, Pete Marsico for technical editing, and finally Steve McGregory and Chris Moore for supporting this research.

\bibliographystyle{IEEEtran}
\bibliography{sparrow}
\end{document}